\documentclass{article} % For LaTeX2e
\usepackage{colm2024_conference}
\colmfinalcopy

\usepackage{microtype}
\usepackage{hyperref}
\usepackage{url}
\usepackage{booktabs}

\title{GenX: Mastering Code and Test Generation with Execution Feedback}

\author{Nan Wang, Yafei Liu, Chen Chen, Haonan Lu \\
OPPO AI Center\\
\texttt{\{wangnan,liuyafei,chenchen4,luhaonan\}@oppo.com} 
}

\usepackage{algorithm}
\usepackage{algorithmic}
\usepackage{amsmath}
\usepackage{booktabs}
\usepackage{multirow}
\usepackage{subcaption}
\usepackage{graphicx}
\begin{document}

\maketitle

\begin{abstract}
Recent advancements in language modeling have enabled the translation of natural language into code, and the use of execution feedback to improve code generation. However, these methods often rely heavily on pre-existing test cases, which may not always be available or comprehensive. In this work, we propose a novel approach that concurrently trains a code generation model and a test generation model, utilizing execution feedback to refine and enhance the performance of both. We introduce two strategies for test and code data augmentation and a new scoring function for code and test ranking. We experiment on the APPS dataset and demonstrate that our approach can effectively generate and augment test cases, filter and synthesize correct code solutions, and rank the quality of generated code and tests. The results demonstrate that our models, when iteratively trained with an increasing number of test cases and code solutions, outperform those trained on the original dataset.
\end{abstract}

\section{Introduction}
Code is a set of instructions that humans write for computers to perform specific tasks, typically crafted by professional programmers using specific programming languages. Assessing the correctness of code is challenging; it usually requires extensive testing to verify whether the program's output aligns with expectations for given inputs. Writing and testing code are both time-consuming tasks. Unlike programming languages, most people are fluent in their natural language. Utilizing artificial intelligence technologies to translate natural language into code and tests holds significant importance in reducing the barriers to programming and enhancing programming efficiency.
\par
Recently, large language models trained on code corpora have demonstrated powerful code generation capabilities. Since code is executable, many efforts have utilized the feedback from execution results to further improve the quality of code generation. One such method involves using reinforcement learning to fine-tune code language models. These approaches \citep{le2022coderl,shojaee2023execution,Liu2023RLTFRL} typically define rewards based on the outcomes of executing the generated code on unit tests (e.g., -0.3 for failing any unit test, 1 for passing all unit tests), and employ algorithms like REINFORCE \citep{schulman2017proximal} or PPO \citep{schulman2017proximal} to optimize the training objective. Another approach is to use execution feedback for better code selection. \citet{li2022competition} uses example tests for filtering and test outputs for clustering. \citet{chen2022codet} utilizes pre-trained language model to generate code and tests directly, and then calculates the dual execution agreement score.
\par
An important prerequisite for using execution feedback is the need for test cases. The APPS \citep{hendrycks2021measuring} dataset is frequently used in many reinforcement learning-based methods for code generation, because it provides annotated test cases in the training set. The well-known AlphaCode \citep{li2022competition} which has achieved human-level performance in competition-level programming, is limited to solving problems that are similar to Codeforces because it heavily relies on example tests provided in the problem description to filter out a large number (99\%) of incorrect solutions. However, many problems in the APPS training set only have a few test cases. Using only these few test cases to judge the correctness of the generated code solution may lead to false positives. Furthermore, the example tests provided in problems from Codeforces do not exist in more realistic code generation scenarios. Many studies on execution feedback for code generation overlook the importance of test generation. To better utilize execution feedback in real-world scenarios for code generation, studying the test generation problem is also crucial.
\begin{figure}
    \centering
    \includegraphics{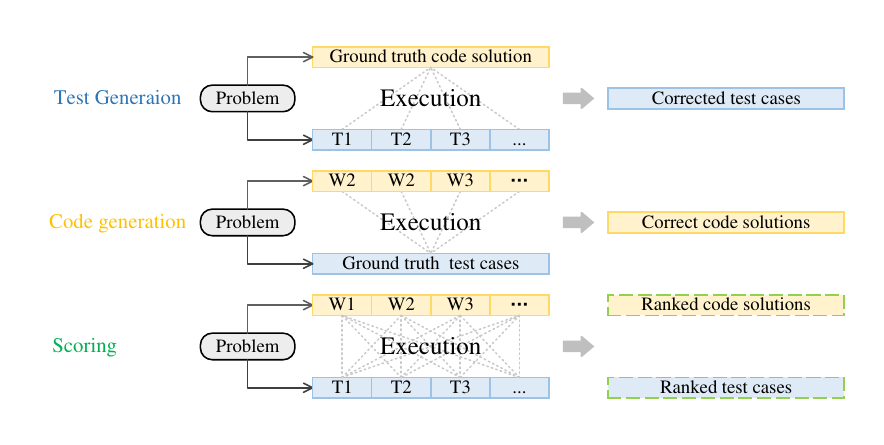}
    \caption{The execution roles under different situations, where W represents code solution, and T represents test case. When there is a ground truth code solution, we can correct the wrong output of the generated tests. When there are ground truth test cases available, we can use them to filter out incorrect code solutions generated and synthesize more correct ones. When only generated code and tests exist, we can execute them against each other for ranking their quality.}
    \label{fig:exec}
\end{figure}
\par
In this work, we train both test generation and code generation models and explore the use of execution feedback in three scenarios. As shown in Figure \ref{fig:exec}, where W and T represent the generated code and test respectively. When we have a ground truth code solution, we can correct the erroneous predictions of the test generation model with it. The corrected test cases can be treated as ground truth test cases. Leveraging this property, we can synthesize more test cases for specific problems. These corrected test cases can be further used to train the test generation model, thereby improving its capabilities. When we have sufficient ground truth test cases, we can use them to judge the correctness of new code generated by the code generation model, thereby avoiding the issue of false positives. By filtering out incorrect code solutions and keeping correct ones, these newly generated correct code solutions can be used to train the code generation model, further enhancing its capabilities. When we lack ground truth for both code solutions and test cases in real scenarios during inference, we can still evaluate the quality of the generated code based on their mutual passing status.
\par
We conducted experiments on the APPS dataset, beginning with the smallest subset of the training set, which included a minimal amount of code and selected tests from the APPS training set. We then incrementally increased the number of test cases and subsequently expanded the code solutions. During the iterative process, the model's code generation and test generation capabilities gradually improved. Ultimately, the performance of the model surpassed that of the model trained on the original APPS dataset. Experiments on the scoring function also demonstrated that our scoring function can effectively filter out high-quality code and tests.
\par
In summary, our contributions are twofold: (1) a framework for simultaneous enhancement of code and test data, expanding the training dataset and resulting in models with superior code and test generation abilities compared to those trained on the initial dataset; (2) a novel scoring function that effectively ranks generated code and tests during inference, filtering for higher quality results..
\section{Related Work}
\paragraph{Code Generation}
In the early stages, many efforts in code generation primarily focused on synthesizing domain-specific languages \citep{Devlin2017RobustFillNP,Bunel2018LeveragingGA} based on input-output examples \citep{Menon2013AML}. As deep learning technologies continued to evolve, the focus shifted towards generating general-purpose code from textual descriptions. CONCODE \citep{2018MappingLT} is one of the early datasets for this purpose. HumanEval \citep{chen2021evaluating}, the most widely evaluated dataset, has been translated into other languages \citep{Zheng2023CodeGeeXAP, Cassano2023MultiPLEAS}. APPS \citep{hendrycks2021measuring} presents a more challenging dataset, comprising three different levels of difficulty. CodeContests \citep{li2022competition} is a more challenging dataset focused on competition-level programming. Unlike the subjective nature of evaluating natural language correctness, the accuracy of code generation is objectively determined by passing extensive test cases.

\paragraph{Test Generation}
Automatic test generation is an important area within software engineering and has been the subject of extensive research \citep{Fraser2011EvoSuiteAT,Fraser2014ALE,Selakovic2018TestGF}. In this work, our focus is primarily on the generation of tests for programming problems. The tests for these programming problems rely solely on the semantics of the problem and can be abstracted as combinations of inputs and outputs. \citet{li2022competition} generate additional test cases by mutating existing test inputs. \citet{chen2022codet,Schfer2023AnEE} leverage the use of code language models to directly generate test cases through prompts. \citet{gorinski2023automatic} use heuristic methods to generate a substantial number of unit tests.

\paragraph{Execution Feedback}
A key characteristic that sets programming code apart from natural language is its ability to be executed by machines. The feedback obtained from executing code can be broadly categorized into two types: textual feedback, such as detailed error messages generated by compilers or interpreters, and binary correctness feedback, such as pass/fail status provided by unit testing  \citep{yang2024if}. Various methods have been developed to leverage these feedback types to enhance code generation abilities. \citet{zhang2023self, chen2023teaching}  utilizes error messages from code execution to refine the original code. \citet{le2022coderl,Shojaee2023ExecutionbasedCG, Liu2023RLTFRL} designs rewards based on unit test outcomes, formalizing code generation as a reinforcement learning problem. \citet{li2022competition}  clusters code based on its output under test inputs and ranks it by cluster size. \cite{chen2022codet} leverages existing pre-trained language models to generate code and test cases, then sorts them according to the agreement in dual execution outcomes.

\paragraph{Data Augmentation}
Training on high-quality data synthesized by large language models can enhance model capabilities \citep{wang2022self, gunasekar2023textbooks, yang2024qwen2}. In code generation, \citet{roziere2023code} produce multiple solutions and unit tests, selecting the first passing solution. \citet{dubey2024llama} use iterative self-correction to refine code and tests. \citet{hui2024qwen2} employ a multilingual sandbox for verification. Unlike approaches that treat test generation as secondary, our work focuses on iteratively enhancing both code and test generation, producing many tests. We retain multiple solutions for fine-tuning to create a more robust dataset.

\section{Approach}
\subsection{Preliminaries}
Code generation, or program synthesis, is a sequence-to-sequence task. Given a problem description $\boldsymbol{p}$, the goal of code generation is to synthesize a program $\boldsymbol{w}=(w_1,w_2,...,w_n)$ that can solve the problem, where each $w_i$ is from a vocabulary and $n$ is the length of the code. Similar to many natural language generation tasks, for a given problem $\boldsymbol{p}$, the exact sequence of $\boldsymbol{w}$ is not unique and may have multiple correct code solutions. 
\par
Test generation refers to the process of creating test cases $T_1, T_2, T_3, ..., T_N$ based on a problem description $\boldsymbol{p}$, where $N$ is the total number of test cases. Each test case $T_i = (x_i, y_i)$ comprises an input $x_i$ and an output $y_i$. 
%For any two tests $(x_i, y_i)$ and $(x_j, y_j)$, if $x_i$ equals $x_j$, then $y_i$ must equal $y_j$. Thus, the number of unique test inputs determines the number of tests. 
The test output is determined solely by the problem definition and the test input, and it is independent of the specifics of the code implementation.  Unlike code generation, where finding a correct code solution is enough, test generation necessitates multiple diverse test cases to cover a variety of scenarios. 
\par
To train a code generation model and a test generation model, we require a training set that meets specific criteria. Each problem in the training set must have at least one associated annotated code solution, a fundamental prerequisite for supervised code generation training. The criteria for test cases are less stringent, with only a subset of the problems necessitating accompanying test cases.
\subsection{Overview}
\begin{algorithm}[!t]
\caption{The overall framework for code and test data augmentation.}
\begin{algorithmic}[1]
\REQUIRE{Seed dataset $D_{\text{seed}}$, pretrained code LLM $M$, iterations $N_t$ and $N_c$}
\ENSURE{Augmented dataset $D_{\text{augmented}}$}
\STATE Initialize $D_{\text{augmented}} \gets D_{\text{seed}}$
\STATE \# Test cases augmentation
\FOR{$i \gets 1$ \textbf{to} $N_t$}
    \STATE Train model $M_t$ on dataset $D_{\text{augmented}}$
    \FORALL{each problem $\boldsymbol{p}$ in $D_{\text{augmented}}$}
        \STATE Sample test cases for $\boldsymbol{p}$ using model $M_t$
        \STATE Parse test inputs and then remove duplicate test cases 
        \STATE Execute all test inputs on the ground truth code solution to obtain outputs
        \STATE Construct new test cases by concatenating test inputs and executed outputs
        \STATE Append valid test cases to the dataset $D_{\text{augmented}}$
    \ENDFOR
\ENDFOR
\vspace{2pt}
\STATE \# Code solutions augmentation
\FOR{$i \gets 1$ \textbf{to} $N_c$}
    \STATE Train model $M_c$ on dataset $D_{\text{augmented}}$
    \FOR{each problem $\boldsymbol{p}$ in $D_{\text{augmented}}$}
        \STATE Sample code solutions for $\boldsymbol{p}$ using model $M_c$
        \STATE Normalize code solutions and then remove duplicate code solutions
        \STATE Execute the generated code solutions on all existing test cases in the dataset $D_{\text{augmented}}$
        \STATE Append valid code solutions to the dataset $D_{\text{augmented}}$
    \ENDFOR
\ENDFOR
\RETURN $D_{\text{augmented}}$
\end{algorithmic}
\label{alg:overall}
\end{algorithm}

We adopt a method similar to self-training \citep{scudder1965probability} to enhance the model's code generation and test generation capabilities. The core of this approach lies in data augmentation by the model itself, which involves obtaining more annotated code solutions and test case annotations. As we trained the test generation model in addition to a code generation model, we make use of the property that they can evaluate the correctness of each other and propose a new scoring algorithm for generated code and tests.
\par
The overall data augmentation framework is depicted in Algorithm \ref{alg:overall}, which comprises two stages. In the first stage, we iteratively augment the dataset with new test cases, requiring only one ground truth code solution. We utilize the test generation model to generate test inputs, and then execute these inputs on the ground truth code solution to obtain ground truth test outputs, thereby obtaining augmented test cases. After acquiring a sufficient number of test cases, we can use them to accurately evaluate the correctness of the code solution. In the second stage, we augment the dataset with additional code solutions. This is achieved by employing iterative rejection sampling on the generated code solutions, discarding incorrect ones, and thereby obtaining a larger set of correct code solutions. In the first stage, using the ground truth code solution to obtain augmented tests ensures no false positives. The extensive test cases help prevent false positives in the second stage of code solution augmentation.
\par
On the augmented dataset, we trained our final code generation and test generation models. During inference, we utilized these models to generate a set of code candidates and a set of test candidates for each problem. These candidates were then executed against each other. Based on the execution results of the code and tests, we assigned a score ranging from 0 to 1 to each generated code and test, and then ranked them accordingly. The specific scoring algorithm is detailed in \ref{alg:score}. Note that the scoring function is used only for ranking during inference, while strict filtering is applied during training data augmentation.
\subsection{Execution Guided Test Generation}
We approach test generation as a sequence-to-sequence task. In this context, the source sequence is the problem description, and the target sequence is a concatenation of multiple test cases. The problem and tests are separated by the string $\langle test \rangle$. For each test case $T_i = (x_i, y_i)$, we convert it into JSON representation of $\{input: x_i, output: y_i\}$. Different test cases are also separated by the string $\langle test \rangle$. The test cases used for training are sourced from both the initial annotations and those added in the last iteration. Since a single problem can have numerous test cases, we randomly select up to 10 for each training step, ensuring a uniform distribution. These selected test cases are then sorted in ascending order by length, as this ordering yields slightly better performance than a random order. The training objective is to minimize the cross-entropy loss on the target sequence.
\par
In each iteration when constructing new test cases, for each problem, we use the model trained on the dataset obtained in the last iteration to sample up to 10 times. Each sampling can yield up to 10 test cases, leading to a potential total of 100 test cases. Each test case is parsed to get the test input, and each test input is executed on the ground truth code solution. Test inputs that fail to execute or have empty outputs are discarded, while those that successfully execute and produce outputs are retained. These successful test inputs, along with the test outputs obtained from execution, serve as the new test cases for the problem. Due to the potential length of test outputs, only those test cases that fall below a specific length threshold are used to train the test generation model in the next iteration. However, they are retained for evaluating code solutions generated in the second stage of data augmentation. To ensure diversity, we remove duplicate test cases.
\par
The test cases we obtain serve two purposes. First, they increase the number of test cases in the training set, which is important to avoid false positives when augmenting code solutions. Second, they are used to train the test generation model. If the goal is simply to augment the number of test cases in the training set, a simpler strategy might be to train a model to generate test inputs. However, we choose to train the test generation model. This is because it already includes the task of generating test inputs and aligns with the model we ultimately want for inference. More importantly, some test cases, obtained by executing test inputs on the ground truth code solution, may not be suitable for training the test generation model. This could be due to them being too long or too trivial (e.g., output is -1).
\subsection{Reject Sampling Code Generation}
During the test case generation phase, we use the execution results of the ground truth code to correct prediction errors made by the model in the test output. However, such strong feedback is not available during code generation. Hence, we employ a rejection sampling method to discard incorrect code solutions produced by the model, retaining only the correct ones. Thanks to the large number of test cases we have gathered for the problem during the previous test data augmentation phase, we can confidently judge the correctness of the newly generated code and avoid false positives.
\par
In the training process of the code generation model, we treat the problem as the source sequence and the correct code solution as the target sequence. These are separated by the string $\left\langle solution \right\rangle$. At each training step, we randomly select a correct code  from all correct solutions available for the problem. The loss function used for training is the cross-entropy loss function applied to the code sequence.
\par
To ensure that correct code solutions are sampled during the use of rejection sampling, we allocate a larger sampling budget than is used during the test generation phase. Furthermore, we adjust the number of sampling attempts based on the varying difficulty levels of the problems: 40 for introductory, 80 for interview, and 160 for competition. After obtaining a code solution, we remove duplicates to ensure diversity among the code solutions. To prevent identical code solutions with different formatting from being considered unique, we also standardize the code formatting.
\par
Code generation involves tasks where multiple correct solutions are possible. For any specific problem, obtaining just one valid solution is adequate. However, we believe that exposing models to a diverse range of code solutions for the same problem can improve their code generation abilities.
\subsection{Dual Critic Scoring}
\begin{algorithm}[!t]
\caption{Code and test scoring algorithm.}
\begin{algorithmic}[1]
\REQUIRE fail/pass matrix $P \in \{0,1\}^{C \times T}$,  iteration times $n$
\ENSURE code scores, test scores
\STATE initialize $code\_scores \in \mathbb{R}^C$ to 1
\STATE initialize $test\_scores \in \mathbb{R}^T$ to 1
\FOR{$i = 1$ to $n$}
    \STATE $code\_scores \leftarrow P \cdot test\_scores / (sum(test\_scores) + 10^{-8})$
    \STATE $test\_scores \leftarrow P^T \cdot code\_scores / (sum(code\_scores) + 10^{-8})$
\ENDFOR
\RETURN $code\_scores, test\_scores$
\end{algorithmic}
\label{alg:score}
\end{algorithm}
In our previous code and test case data augmentation framework, we used the ground truth code solution to evaluate the generated test cases and used the ground truth test cases to evaluate the generated code solutions. However, this approach is not feasible in practice, as the ground truth code solutions are not available during inference and can only be used for the training set. To address this limitation, we propose a scoring algorithm that does not depend on the ground truth code solutions. Instead, it only requires the pass/fail status of the generated test cases when run against the code solutions.
\par
Algorithm \ref{alg:score} outlines the overall scoring procedure. Initially, we are provided with several generated code solutions and corresponding test cases for a specific problem. These test cases are then executed using the code solutions, yielding a fail/pass status for each, i.e., a binary 0/1 matrix $P$. Starting with $P$, we iteratively compute the scores for both the code solutions and the test cases. The score of each code solution is determined by the scores of the test cases it passes, and vice versa. To ensure that the scores remain within the 0 to 1 range during each iteration, we normalize all scores before proceeding to the next iteration. We simulated the convergence of Algorithm \ref{alg:score} under different inputs, and it typically converges within a few hundred iterations.
\par
In this scoring algorithm, $code\_scores$ and $test\_scores$ can be approximately viewed as the proportion of code or tests that successfully pass each other. Assuming all tests are correct, the score computed in the scoring function for the code represents the proportion of test cases that the code passes. A similar scenario applies when all the code is correct. The core of this algorithm is the reciprocal evaluation between code and tests, akin to a dual critic system.
\section{Experiments}
\subsection{Experimental Setup}
\paragraph{Dataset}
\label{app:data}
\begin{table}[!ht]
\centering
\begin{tabular}{@{}lccc@{}}
\toprule
Dataset & Problems & Solutions per problem & Tests per problem \\ \midrule
APPS    & 5000     & 23.45                 & 5.16              \\
APPS-   & 4977     & 1.00                  & 1.38              \\
APPS+   & 4977     & 20.14                 & 184.09            \\ \bottomrule
\end{tabular}
\caption{Statistics of the number of problems, solutions, and tests on APPS, APPS-, and APPS+.}
\label{tab:stat}
\end{table}
We conducted experiments on the \textbf{APPS} dataset, which includes programming problems of three difficulty levels: introductory, interview, and competition. To demonstrate the effectiveness of our method, we removed a significant portion of the code and tests from the APPS data, creating a new \textbf{APPS-} dataset. The process of constructing the APPS- training set is as follows: First, we removed all code solutions with syntax errors, kept correct code solutions, and formatted them using black formatter \citep{Langa_Black_The_uncompromising}. We then retained only the shortest code solution. We also eliminated problems that did not have any solutions. For test cases, we retained the shortest 10 if a problem had more than 10; otherwise, we discarded all of them. The final APPS- training set contains 4977 problems, each with a corresponding code solution. Of these, 686 problems have 10 test cases, while others have none. To construct the APPS- test dataset, we filtered out problems in the APPS test set that did not have valid code solutions. Then, we uniformly sampled 100 test cases from each difficulty level, resulting in a final test set size of 300. We also only keep the shortest code solution to validate the correctness of test cases. Based on APPS-, the final dataset, which has undergone testing and code data augmentation, is referred to as \textbf{APPS+}. We kept short code and tests due to the limited context window in our training, as well as for simplicity. Detailed statistics of these datasets are presented in Table \ref{tab:stat}. The number of problems in APPS- and APPS+ is fewer than 5000 because those with syntax errors have been excluded. Notably, in APPS-, 4291 problems lack test cases, which results in fewer tests per problem.
\paragraph{Evaluation} 
For code generation, consistent with previous work \citep{chen2021evaluating, le2022coderl, li2022competition}, we use pass@k as the metric when no scoring is applied to the code, and n@k as the metric after scoring the code. The calculation of n@k involves selecting the top-n scored code solutions (which may exceed n) and then calculating their pass@k. For test generation, no widely accepted evaluation metrics exist. As diversity is crucial for test generation, we first remove duplicates from the generated tests, then propose two metrics to evaluate the unscored generated tests: pass rate, which represents the percentage of generated test cases that pass the ground truth code solution, and pass num, which denotes the number of passed test cases. When evaluating scored generated test cases, we propose a new metric, \textbf{Pr@n}, which represents the average pass rate of the top-n scored test cases.
\paragraph{Implementation Details}
We conducted our experiments on DeepSeek Coder 5.7b \citep{guo2024deepseek}, as it utilizes multi-query attention \citep{shazeer2019fast} which enables faster inference speeds, an important factor in our approach. The number of iterations for code and test generation was set to 3. To get more test cases, we also conducted additional test sampling after the third iteration. For training, we used a linear warmup for 100 steps, then reached a maximum learning rate of 1e-5, followed by a cosine decay to 1e-6, and trained for three epochs. We used a maximum length of 1536 tokens for the source sequence and 1024 tokens for the target sequence. During inference on the test set, we used nucleus sampling with top p=0.95 and a temperature of 0.8 for test generation, and a temperature of 0.6 for code generation. When sampling on the training set for data augmentation, we raised the temperature setting by 0.2. The code execution was carried out on Python version 3.8 and the operating system was Ubuntu 20.04.
\subsection{Main Results}
\begin{table}[]
\begin{minipage}{0.5\textwidth}
\begin{tabular}{lcccccc}
\toprule
\multirow{2}{*}{Dataset} & \multicolumn{3}{c}{Pass rate} & \multicolumn{3}{c}{Pass num} \\ 
\cmidrule(lr){2-4} \cmidrule(lr){5-7} 
                         & Intro    & Inter    & Comp    & Intro    & Inter    & Comp   \\ 
\midrule
APPS  & 30.34    & 24.66    & 23.85   &  12.99	& 14.08	 &  12.82  \\
APPS- & 31.97    & 27.37    & 27.50   & 20.99     & 18.1     & 17.66   \\
APPS+ & \textbf{34.10}    & \textbf{29.77}    & \textbf{26.93}   & 27.49     & 25.96     & 23.97   \\ 
\bottomrule
\end{tabular}
\end{minipage}
\begin{minipage}{0.5\textwidth}
\flushright
\vspace{16pt}
\begin{tabular}{lcc}
\toprule
Difficulty & Pr@1 & Pr@10 \\ 
\midrule
Intro      &  72.61            &   62.13          \\
Inter      &  65.74            &   53.78          \\
Comp       &  64.21            &   52.36          \\ 
\bottomrule
\end{tabular}
\end{minipage}
\caption{Pass rate (\%) and pass num for unranked test generation results on three datasets (left), and Pr@1 (\%) and Pr@10 (\%) for ranked test generation results on the APPS+ dataset (right).}
\label{tab:main_test}
\end{table}
\begin{table}[!t]
\begin{minipage}{0.5\textwidth}
\begin{tabular}{lcccccc}
\toprule
\multirow{2}{*}{Dataset} & \multicolumn{3}{c}{Pass@1} & \multicolumn{3}{c}{Pass@10} \\
\cmidrule(lr){2-4} \cmidrule(lr){5-7}
                          & Intro    & Inter   & Comp   & Intro & Inter & Comp                        \\
\midrule
APPS     & 21.90    & 1.50    & 0.30   & \textbf{42.00} & 7.00 & 3.00 \\
APPS-    & 20.90    & 2.10    & 0.60   & 40.00 & \textbf{8.00}     & \textbf{5.00}                        \\
APPS+    & \textbf{23.40}    & \textbf{2.20}    & \textbf{1.10}   & 38.00 & $7.00$                     & 4.00                        \\
\bottomrule
\end{tabular}
\end{minipage}
\begin{minipage}{0.5\textwidth}
\flushright
\vspace{16pt}
\begin{tabular}{lcc}
\toprule
Difficulty & Pass@1 & 1@10 \\ 
\midrule
Intro      &   23.40   &   \textbf{30.60}     \\
Inter      &   2.20    &    \textbf{4.00}    \\
Comp       &   1.10    &    \textbf{2.00}    \\ 
\bottomrule
\end{tabular}
\hspace{0pt}
\end{minipage}
\caption{Pass@k (\%) for unranked code generation results on three datasets (left), and n@k (\%) for ranked code generation results on the APPS+ dataset (right).}
\label{tab:main_sol}
\end{table}
In Table \ref{tab:main_test}, we present the results of test generation on the three different datasets. For each problem, we performed 10 samplings, thus generating a total of up to 100 test cases across all problems. The results indicate that the model trained on the APPS+ dataset outperforms the one trained on the APPS- dataset, both in terms of pass rate and pass num, with the improvement in pass num being particularly significant. It is noteworthy that the original APPS dataset performs quite poorly in terms of the pass num metric, even worse than APPS-, which is only a small subset of APPS. This may be due to the uneven distribution of the number of test cases in the full APPS dataset, where a large number of problems only contain one test case, preventing the model from generating a full set of 10 test cases. 
\par
Table \ref{tab:main_sol} shows the code generation results on the three datasets. We generated 10 solutions for each problem and evaluated two metrics: pass@1 and pass@10. In terms of code generation, the pass@10 metric reflects the diversity of the model. As in practical applications, one correct code solution is enough, so the pass@1 metric is more important. The results show that the model trained on the APPS+ dataset performs better on the pass@1 metric, even surpassing the model trained on the complete APPS dataset. However, this model scores lower on the pass@10 metric, which may be due to the lack of diversity in the augmented data. It is also worth noting that the performance of code generation is significantly poorer at interview and competition levels than at the introductory level. This is in contrast to test generation, which only experiences a slight drop in performance as difficulty increases. This implies that using a test generation model to predict test output is easier than generating correct code.
\par
The ranked results in Table \ref{tab:main_test} and Table \ref{tab:main_sol} further validate the effectiveness of our scoring function. Selecting high-scoring code solutions or test cases is better than random selection. Even under interview and competition difficulty levels, where the pass@1 metric for generated code is very low, the use of the scoring function still achieves significant improvement. We also evaluated on HumanEval and MBPP in Appendix \ref{full_eval}.
\par
\subsubsection{Comparison with CodeT}
\label{app:codet}
\begin{table}[]
\centering
\begin{tabular}{@{}lllll@{}}
\toprule
\multirow{2}{*}{Difficulty} & \multicolumn{2}{c}{CodeT}                                   & \multicolumn{2}{c}{GenX}                                    \\ \cmidrule(lr){2-3} \cmidrule(lr){4-5} 
                            & \multicolumn{1}{c}{1@10} & \multicolumn{1}{c}{Pr@10} & \multicolumn{1}{c}{1@10} & \multicolumn{1}{c}{Pr@10} \\ 
                    \midrule
Intro                       & 29.60                 & 52.06                         & \textbf{30.60}         & \textbf{62.13}                 \\
Inter                       & 3.87            & 40.75                        & \textbf{4.00}          & \textbf{53.78}                 \\
Comp                        & \textbf{2.00}          & 39.40                          & \textbf{2.00}          & \textbf{52.36}                 \\ \bottomrule
\end{tabular}
\caption{Comparison of 1@10 (\%) for code generation and Pr@10 (\%) for test generation between CodeT and our method GenX.}
\label{tab:codet}
\end{table}
Our scoring algorithm is akin to that of CodeT \citep{chen2022codet}, as both are predicated on the dual execution of code and tests. CodeT computes scores by identifying a consensus set and evaluating code solutions based on this set, with the potential scores being any positive integer. In contrast, our method calculates scores akin to the pass rate between code solutions and test cases, yielding scores that range from 0 to 1. For comparison purposes, we have replicated CodeT's scoring function as described in their paper, assigning uniform scores to all code and tests within a consensus set. As indicated in Table \ref{tab:codet}, our scoring function slightly outperforms CodeT in terms of code generation. In the realm of test generation, our function's scores are significantly higher than those of CodeT, which may be attributed to the fact that the consensus set approach is more suited to evaluating code than tests.

\subsection{Analysis}
\paragraph{Impact of False Positives}
\begin{table}[]
\centering
\begin{tabular}{@{}lcccccc@{}}
\toprule
\multirow{2}{*}{Dataset} & \multicolumn{3}{c}{Pass@1} & \multicolumn{3}{c}{Pass@10} \\ 
\cmidrule(lr){2-4} \cmidrule(lr){5-7} 
                           & Intro   & Inter   & Comp   & Intro    & Inter   & Comp   \\ \midrule
APPS                       & 20.90   & 1.70    & 0.60   & 36.00    & 7.00    & 3.00   \\
APPS+                      & \textbf{21.50}   & \textbf{3.20}    & \textbf{0.80}   & \textbf{40.00}    & \textbf{7.00}    & \textbf{4.00}   \\ 
\bottomrule
\end{tabular}
\caption{Experimental results of pass@k (\%) for code generation using one rejection sampling iteration, where APPS has fewer test cases compared to APPS+.}
\label{tab:fp}
\end{table}
Before implementing rejection sampling to iteratively augument code solutions, we performed several rounds of test data augmentation. These data were instrumental not only in training the final test generation model but also in evaluating the generated code. The goal was to minimize the introduction of false positives in code solutions due to an insufficient number of tests, which could have caused the model to learn incorrect coding patterns. As shown in Table \ref{tab:fp}, utilizing test cases from APPS+ led to improved code generation performance. Although use APPS retained more code solutions, its limited number of test cases resulted in a higher incidence of false positives.

\paragraph{Influence of Replay}
\begin{figure}[htbp]
\centering
\begin{subfigure}{0.45\textwidth}
\includegraphics[width=\textwidth]{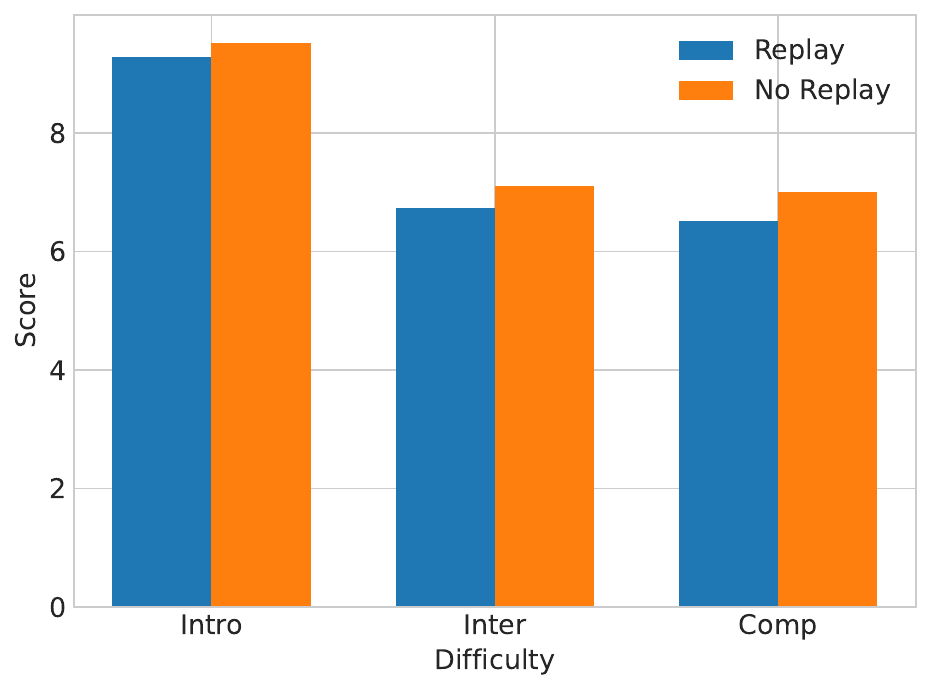}
\caption{Test generation results.}
\end{subfigure}
\hfill
\begin{subfigure}{0.45\textwidth}
\includegraphics[width=\textwidth]{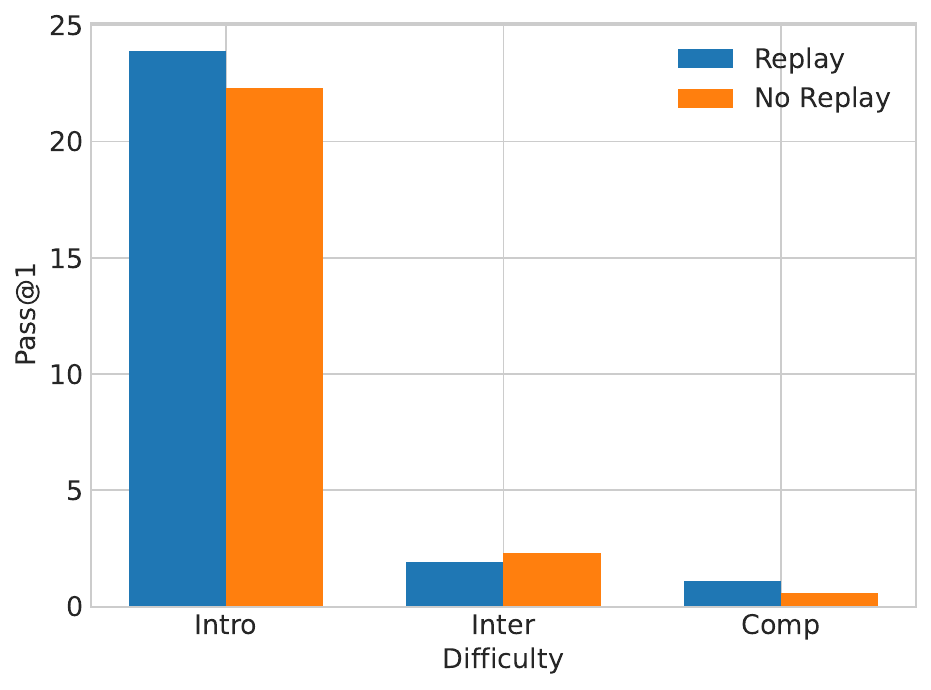}
\caption{Code generation results.}
\end{subfigure}
\caption{The experimental results of test generation and code generation when replaying historical samples during the second iteration. Score is defined as the product of pass rate and pass num.}
\label{fig:replay}
\end{figure}
In the iterative augmentation of code and test data, each iteration requires training the corresponding model. The training data can consists of either all the data generated from previous iterations (replay) or just the data produced by the last iteration's model (no replay). We have found that the decision to use replay varies between code generation and test generation. As Figure \ref{fig:replay} shows, for test generation, avoiding replay is the most effective strategy. However, for code generation, incorporating previous data leads to better outcomes. This discrepancy may arise from the dataset size we can produce. For one sampling time, we can generate ten test cases, while we can only generate one code solution. Furthermore, test cases with errors can be corrected; erroneous code must be discarded.

\paragraph{Improvements During Iterations}
\begin{figure}
\centering
\begin{subfigure}{0.45\textwidth}
\includegraphics[width=\textwidth]{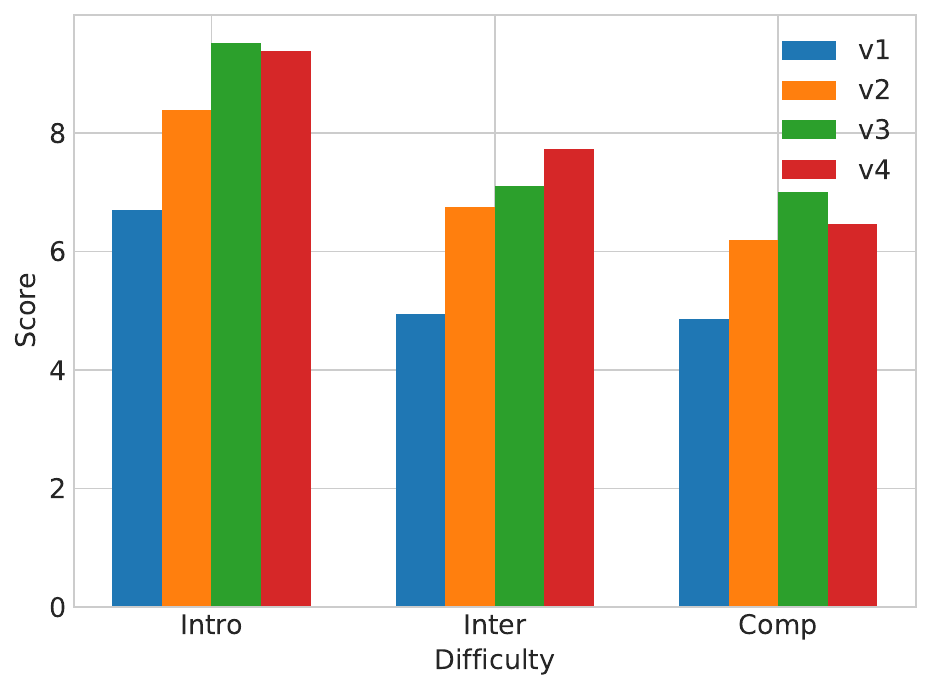}
\caption{Test generation results.}
\end{subfigure}
\hfill
\begin{subfigure}{0.45\textwidth}
\includegraphics[width=\textwidth]{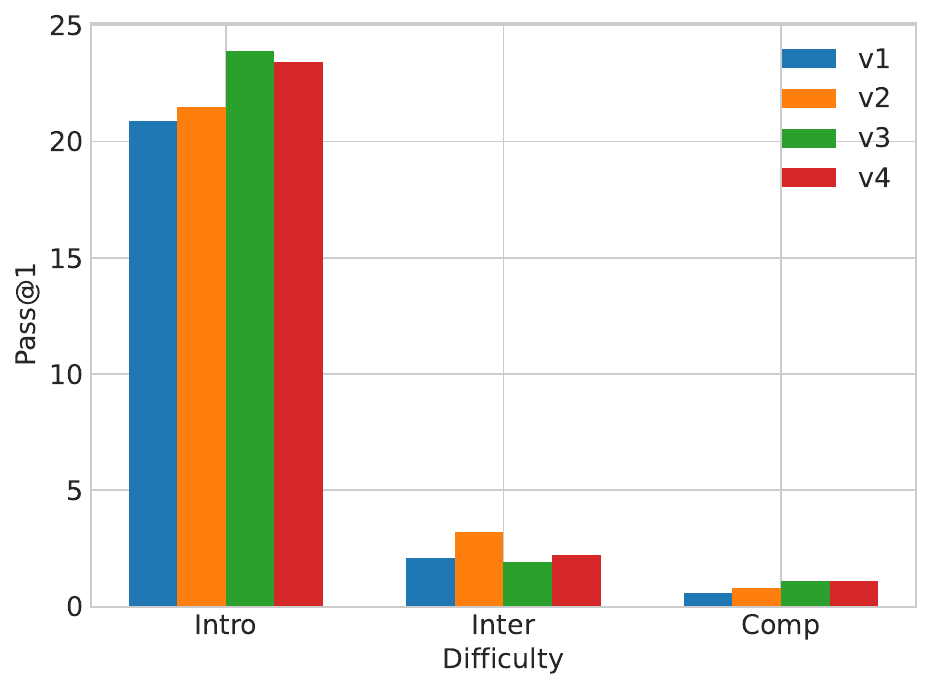}
\caption{Code generation results.}
\end{subfigure}
\caption{The experimental results under three iterations and four different versions (v1 to v4) of data for test generation and code generation. Score is defined as the product of pass rate and pass num.}
\label{fig:version}
\end{figure}
In Figure \ref{fig:version}, we have plotted the performance changes of the code generation model and the test generation model throughout the iterative process. It can be observed that both models experienced an improvement in performance from version 1 to version 2 during the first iteration. However, in subsequent iterations, the performance varied under problems of different difficulty levels. For the test generation model, the performance on interview level problems improved consistently, but there was a decline in performance on introductory and competition level problems after the third iteration. As for the code generation model, the performance on introductory problems peaked during the second iteration, while the performance on interview level problems was highest during the first iteration, and the performance on competition level problems stopped improving after the third iteration.
\section{Conclusion}
In this work, we investigate the interaction between code and test cases in three different scenarios: when the code serves as the ground truth, when the test cases are considered correct, and when both code and test cases are generated. In the first scenario, we use the execution results to correct errors in predictions made by the test model. In the second, we employ rejection sampling to ensure the retention of only those code solutions that are correct. In the third scenario, we design a scoring function that evaluates the quality of both the code and the test cases based on their interaction. 
\par
Currently, We do not utilize additional information, such as text-rich error messages, which could potentially be beneficial for repairing code that is discarded during the rejection sampling process. We leave this as future work.

\bibliography{colm2024_conference}
\bibliographystyle{colm2024_conference}

\appendix
\section{Appendix}
\label{full_eval}
\subsection{Evaluation on HumanEval and MBPP}

\begin{table}[h]
\centering
\small
\begin{tabular}{lcccccc}
\toprule
Test Set & Model & Pass@1 (\%) & Pass@10 (\%) & Pass Rate (\%) & Pass Num \\
\midrule
\multirow{3}{*}{HumanEval} 
& GenX - APPS- & 39.03 & 67.74 & 48.12 & 29.66 \\
& GenX - APPS+ & 45.61 & 72.90 & 46.39 & 36.99 \\
& GPT-3.5 Turbo & 72.90 & 90.32 & 47.55 & 24.85 \\
\midrule
\multirow{3}{*}{MBPP} 
& GenX - APPS- & 41.27 & 61.65 & 30.86 & 20.77 \\
& GenX - APPS+ & 44.72 & 62.05 & 32.05 & 26.10 \\
& GPT-3.5 Turbo & 53.33 & 66.06 & 28.30 & 17.39 \\
\bottomrule
\end{tabular}
\caption{Code generation and test generation results on HumanEval and MBPP test sets.}
\label{tab:direct_eval}
\end{table}

\begin{table}[h]
\centering
\small
\begin{tabular}{lccccc} 
\toprule
Test Set & Model & \multicolumn{2}{c}{CodeT} & \multicolumn{2}{c}{GenX} \\ 
\cmidrule(lr){3-4} \cmidrule(lr){5-6}
& & 1@10 & Pr@10 & 1@10 & Pr@10 \\ 
\midrule
\multirow{2}{*}{HumanEval} 
& GenX - APPS+ & 58.56 & 76.42 & \textbf{61.22} & \textbf{82.90} \\
& GPT-3.5 Turbo & 81.99 & 64.81 & \textbf{82.18} & \textbf{67.34} \\ 
\midrule
\multirow{2}{*}{MBPP} 
& GenX - APPS+ & 46.77 & 47.87 & \textbf{47.49} & \textbf{54.33} \\
& GPT-3.5 Turbo & 56.69 & 42.32 & \textbf{57.17} & \textbf{45.02} \\
\bottomrule
\end{tabular}
\caption{1@10 (\%) and Pr@10 (\%) results on HumanEval and MBPP test sets.}
\label{tab:rank_eval}
\end{table}

To further evaluate performance on the HumanEval and MBPP datasets, we converted these datasets into the APPS format and utilized the APPS evaluation script. Experiments were conducted using both the GenX models (trained on APPS- and APPS+) and GPT-3.5 Turbo. Note that due to conversion errors and differences in the APPS evaluation script, the results may not align with official evaluations. However, all models were evaluated under consistent settings to ensure a fair comparison.

Table \ref{tab:direct_eval} presents the results of unsorted code and test generation on the HumanEval and MBPP datasets. Models trained on APPS+ demonstrated improved performance compared to those trained on the initial dataset, APPS-. Notably, in test generation, our models even outperformed GPT-3.5 Turbo, indicating that a smaller, fine-tuned language model can surpass a larger model in direct prompt-based generation.

Table \ref{tab:rank_eval} shows the results of GenX and CodeT after sorting. Similar to the ranking results on APPS in the main results section, GenX outperformed CodeT in both code and test generation, particularly excelling in test generation ranking.
\end{document}